\documentstyle[aps,prb,preprint,epsfig,floats]{revtex}

\begin{document}
\title{Cure to the Landau-Pomeranchuk and Associated  Long-Wavelength 
Fermi-surface Instabilities on Lattices }

\author{C. M. Varma}
\address{Bell Laboratories, Lucent Technologies,  
Murray Hill, NJ 07974, and \\
University of California, Riverside, CA. 92521}
\maketitle

\begin{abstract}

The cure to the $\ell=1$ Landau-Pomeranchuk (L-P) instabilities in translationally invariant
fermions is shown to be a state with an anisotropic gap at the fermi-surface. For higher $\ell$ and for fermions on a 
lattice, general criteria for long wavelength instabilities and their cure are found in terms of
the derivatives of the single particle self-energy with respect to momentum for spin-symmetric 
instabilities and with respect to magnetic field for spin-antisymmetric instabilities. 
The results may be relevant to identifying hidden order parameters found in many metals.

\end{abstract}

\maketitle

 If the equilibrium distribution function $n({\bf k})$
of a translationally invariant (TI)
 Fermi-liquid is changed by  $\delta n({\bf k},\sigma)$,
 \begin{eqnarray}
 \delta n({\bf k},\sigma) = \sum_{\ell} \left[\delta n^s_{\ell}+\delta n^a_{\ell}\right]
 P_{\ell }(\theta) \delta(\epsilon_{\bf k}-\mu),
 \end{eqnarray}
the system is stable only if 
 \begin{eqnarray}
F_{\ell}^{s,a} > -(2{\ell}+1).
\end{eqnarray}
The superscripts $s,a$ refer to the symmetric and
antisymmetric spin-channels and $F_{\ell}^{s,a}$ are the Landau
parameters.
These conditions derived by Pomeranchuk \cite{pomeranchuk}are the
generalizations of the
 positivity conditions
on the compressibility, magnetic susceptibility and the specfic-heat, derived
 by Landau \cite{landau}.

 Suppose the inequalities (3) are violated for some $\ell$.
 What is the symmetry and low-energy excitation spectrum in the state which cures
 the instability? Obviously, rotational symmetry must be lost for $\ell \neq 0$. But
 rotational symmetry can be lost
  in at least two different kinds of
 states with quite different excitation spectra, states with a distorted Fermi-surface
 as is commonly presumed, or by a state with an anisotropic gap at
 the Fermi-surface. Starting from an
 equation of motion for the distribution function and using the relations between interactions and single-particle self-energy (Ward
 identitites), I show
that the loss of the rotational symmetry
 of the distribution function at least for the $\ell= 1$ case must occur through
 a state with an anisotropic gap in the distribution function. 
 For the spin-antisymmetric $\ell=0$ instability, the answers from
 this approach are identical to that in the usual theory of itinerant ferromagnetism. 
 For higher $\ell$ and for fermions on a lattice two classes of spin-symmetric instabilities are identified; one is cured by a state with an anisotropic gap at the fermi-surface, the other by a state with an anisotropic distortion of the fermi-surface. For spin-antisymmetric instabilities, the cure is a state with an anisotropic gap for spin-flip excitations at the fermi-surface.

At the empirical end, there is evidence of
 gaps with "hidden order parameters"\cite{uru2si2,ceal3,cmv} in many materials for which
 this work may be relevant.

Let us specify quasi-particles created and annihilated by
operators $c^+_{{\bf k}\sigma},c_{{\bf k}\sigma}$ respectively
with quasiparticle energy $\epsilon_{\bf k}$ and an interaction
function  $F({\bf k,k',q},\sigma,\sigma')$ whose forward
scattering limit ${\bf q}\rightarrow 0$ is the Landau interaction
function. Specify further that
  $F$ has a cut-off at $\Omega_c$ so that it approaches zero for
   $|\epsilon_{\bf k}-\mu|,|\epsilon_{\bf k'}-\mu| \gtrsim \Omega_c$.
   Attractive
    Landau parameters might require that $F({\bf k,k',q},\sigma,\sigma')$
    come through the exchange of some Bosons with an upper energy cut-off
    $\Omega_c$ generated by the repulsive bare interactions which have a much larger
    cut-off. (See discussion at the end.)

    Consider first the instabilities in the spin-symmetric channel. The equation of motion for
    small ${\bf q}$
    is the {\it homogeneous} Landau transport equation:
\begin{eqnarray}
 {(\epsilon_{{\bf{k}}+
{\bf{q}}/2}-\epsilon_{{\bf{k}}-{\bf{q}}/2}-\omega)}<c^+_{{\bf{k}}+{\bf{q}}/2}
c_{{\bf{k}}-{\bf{q}}/2}>_s = \\ \nonumber
~~~~~~~~~~~[f(\epsilon_{{\bf{k}}+{\bf{q}}/2})
-f(\epsilon_{{\bf{k}}-{\bf{q}}/2})]& \sum_{{\bf{k}}'}
F_s({\bf{k}},{\bf{k}}')
 <c_{{\bf{k}}'+{\bf{q}}/2}^+c_{{\bf{k}}'-{\bf{q}}/2}>_s,
 \end{eqnarray}
 in which $f(x)$ is the Fermi-distribution.

    If the forward scattering amplitude is
 attractive enough for any $\ell$ to violate (2), no solution of Eq. (3)
 is possible for $\omega \geq 0$ in that angular momentum channel.
 The Fermi-liquid is then unstable and must reconstruct. The stability condition
 is given by putting $\omega=0$ and then taking the limit ${\bf q}\rightarrow 0$
 of Eq. (3) and may be written as
 \begin{eqnarray}
  \Phi_s({\bf{k+q/2}},{\bf{k-q/2}}) \geq \sum_{k^{\prime}}^{\prime}
F_s({\bf{k},{k}}')\frac{f(\epsilon_{{\bf{k}}'+{\bf{q}}/2})
-f(\epsilon_{{\bf{k}}'-{\bf{q}}/2})}{(\epsilon_{{\bf{k}}'+{\bf{q}}/2}-
\epsilon_{{\bf{k}}'-{\bf{q}}/2})}
 \Phi_s({\bf{k'+q/2}},{\bf{k'-q/2}}),
\end{eqnarray}
\begin{eqnarray}
[f(\epsilon_{{\bf{k}}})-f(\epsilon_{{\bf{k}}'})]\Phi_s({\bf{k},{k}'})
= (\epsilon_{{\bf{k}}}-\epsilon_{{\bf{k}}'})
<c_{{\bf{k}}}^{\dagger}c_{{\bf{k}}' }>_s.
\end{eqnarray}
The cure for this instability is to find a new set of
quasi-particle operators
 $\tilde{c}^{\dagger}_{{\bf{k}},\sigma},\tilde{c}_{{\bf{k}},\sigma}$, with energies
 $E_{\bf k}$, which are linear combinations of the operators
 $ c^{\dagger}_{{\bf{k}},\sigma}c_{{\bf{k}},\sigma}$ such that the
 stability condition is satisfied.

In the new (stable) state the stability condition is a new Landau
transport equation analogous to Eq. (4), where again the limit
${\bf q}\rightarrow 0$ is taken:
\begin{eqnarray}
\tilde{\Phi}_s({\bf{k+q/2},{k-q/2}}) \geq \sum_{k^{\prime}}'
F({\bf{k},{k}}')\frac{f(E_{{\bf{k}}'+{\bf{q}}/2})
-f(E_{{\bf{k}}'-{\bf{q}}/2})}{(E_{{\bf{k}}'+{\bf{q}}/2}-E_{{\bf{k}}'-{\bf{q}}/2})}
 \tilde{\Phi}_s({\bf{k'+q/2},{k'-q/2}}),
\end{eqnarray}
\begin{eqnarray}
[f(E_{{\bf{k}}})-f(E_{{\bf{k}}'})]\tilde{\Phi}_s({\bf{k},{k}'}) =
(E_{{\bf{k}}}-E_{{\bf{k}}'})
<\tilde{c}_{{\bf{k}}}^{\dagger}\tilde{c}_{{\bf{k}}'}>_s.
\end{eqnarray}
In Eq.(6), the sum is restricted to states
 such that $E_{{\bf k\pm q}}$ are within about $\Omega_c$ of the chemical potential $\tilde{\mu}$.
 It has been assumed that the quasiparticle interactions in the new stable phase
 are not different from those of the unstable phase. This is only true if the
 changes $(E_{\bf k}-\epsilon_{\bf k}) \ll E_f$, i.e. in the weak-coupling limit. Strong-coupling
 corrections are not considered here.

 Replacing $\geq$ by equality, the integral equation (6)
serves as the self-consistency condition to determine the
temperature of the instability and the new quasi-particle energies
$E_{\bf k}$.
 Eq.(6)may also
 be derived in specific models and for fermions on a lattice. One such derivation,
  which also serves to indicate a possible origin
 of the negative Landau parameters, is given at the end.
 
Suppose the instability occurs in the $\ell$-th angular
 momentum channel. Write
  $F^s({\bf k,k}')$ in a separable form and  its $\ell$-th component as 
  $\nu^{-1}(0)F_{\ell}^s P_{\ell}(\cos \theta_k)P_{\ell}(\cos \theta_k')$, where $\nu(0)$ is
 the density of states at the chemical potential. Then the solution of
 Eq. (6) has the form:
 \begin{eqnarray}
 \tilde{\Phi}_s({\bf k})~ & \propto ~P_{\ell}(\cos\theta_k),
\end{eqnarray}
and the the self-consistency condition  in the limit ${\bf{q}} \rightarrow 0$ is
 \begin{eqnarray}
 1 =  F_{\ell}^s\nu(0)^{-1}\sum ^{'}_{{\bf{k}}}P_{\ell}^2(\cos \theta_k)\left(
  \frac{f(E^{>}_{{\bf{k}}})-
  f(E^{<}_{{\bf{k}}})}{E^{>}_{{\bf{k}}}
   -E^{<}_{{\bf{k}}}}\right).
  \end{eqnarray}
Here $E^{>}_{{\bf{k}}},E^{<}_{{\bf{k}}}$ are the new one particle eigenvalues
above and below the chemical potential respectively.

The analogous equation when the instability is in the $\ell$-th
spin-antisymmetric channel is
\begin{eqnarray}
 1 =  F_{\ell}^a\nu(0)^{-1}\sum ^{'}_{{\bf{k}}}P_{\ell}^2(\cos \theta_k)\left(
  \frac{f(E^{>}_{{\bf{k}},\uparrow})-
  f(E^{<}_{{\bf{k}},\downarrow})}{E^{>}_{{\bf{k}},\uparrow}
   -E^{<}_{{\bf{k}},\downarrow}}\right),
  \end{eqnarray}
where $E^{>}_{{\bf{k}},\uparrow},E^{<}_{{\bf{k}},\downarrow}$ are
the new one particle eigenvalues for up and down-spin and above
and below the chemical potential respectively. In the absence of
spin-orbit scattering the spin may be quantized by the helicity,
i.e. along and opposite ${\bf k}$.

  Unlike BCS type theories which have an order parameter in
an {\it orthogonal} channel, a self-energy in quadrature to the
single-particle energy is not expected in the present problem, and I make the ansatz
   \begin{eqnarray}
E_{{\bf{k}}}\equiv \epsilon_{{\bf k}}+D_{{\bf{k}}}.
\end{eqnarray}
This ansatz is shown below to be consistent at least in the region
just below the transition temperature.The form of $D({\bf k})$ is now picked through the microscopic
theory for the approach to the instability. This is most
transparently done for the $\ell = 0$ and $1$ channels.
Consider the
  single-particle self-energy $\Sigma({\bf k},\omega)$ in the stable state as the
  L-P instability is approached. The instability in the symmetric $\ell=1$
  channel is the vanishing of the effective mass $m^*$ given
  by \cite{invariance}
  \begin{eqnarray}
  m^*/m  =   z^{-1}(1+(v_{f0})^{-1}\partial \Sigma/\partial {\bf k}|_{k_F,\mu})^{-1}  
               =   1 + F_{1s}/3.
  \end{eqnarray}
In Eq.(12) $v_{f0}$ is the free particle velocity, and $z$ is the
quasiparticle renormalization
  amplitude, $z=(1-\partial \Sigma
  /\partial \omega)^{-1}$.  The first part of Eq. (12) is always valid; the second part relies on TI. Since $0< z \leq 1$, $m^*/m$
  can be less than 1 only if
   $\Sigma({\bf k}, \epsilon_{\bf k})$
  is negative/positive for ${\bf k}$ below/above ${\bf k}_f$,
  or on the energy shell below/above the chemical
  potential $\mu$.
   This is illustrated in Fig. (1) for the case the self-energy vanishes above $\Omega_c$.
    Curve 1 is the unrenormalized $\epsilon_{\bf k}$, curve 2 and 3 show the changes
   in the quasiparticle dispersion to $\epsilon_{\bf k}+ \Sigma({\bf k}, \epsilon_{\bf k})$,
    as the instability is approached. At the instability,
   the dispersion at the chemical potential is vertical (i.e. $m^*=0$ or divergent velocity) and negative beyond it.
 \begin{figure}[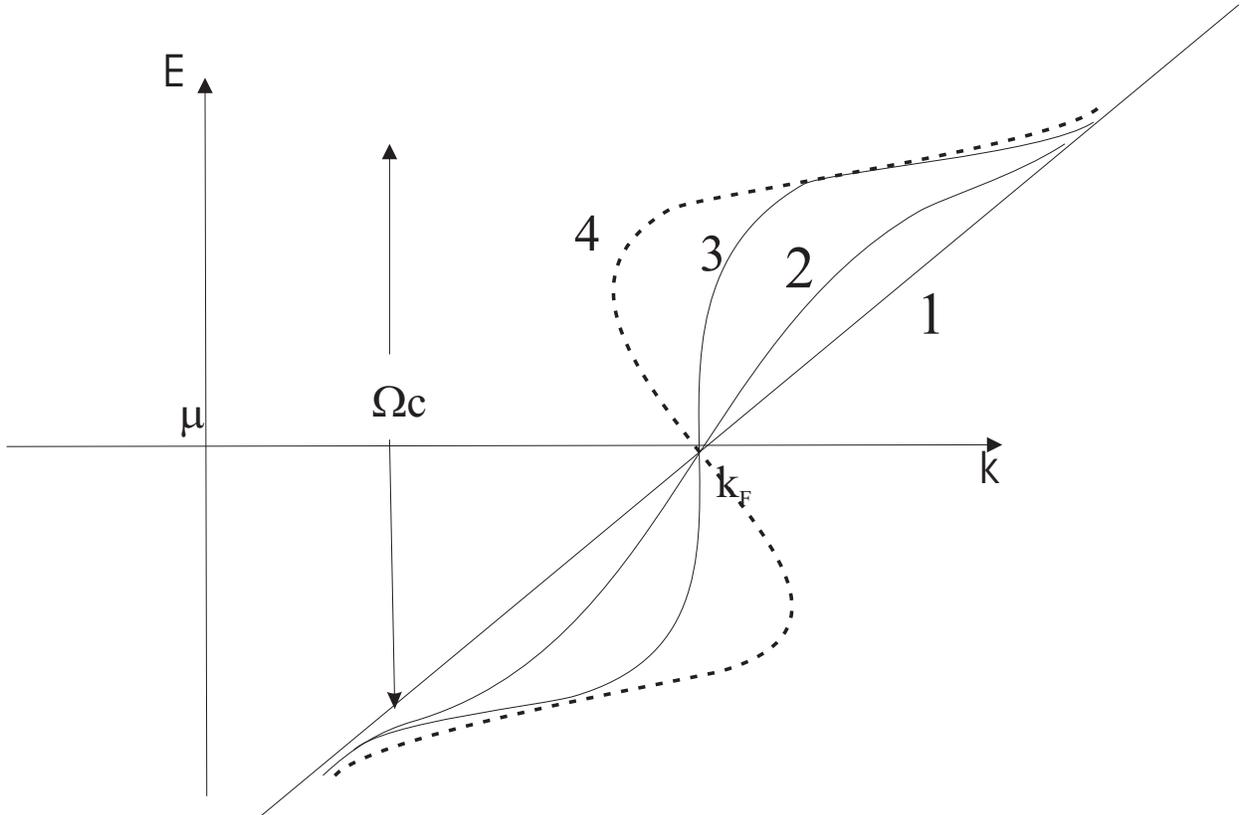]
\centering
\epsfxsize=\linewidth
\epsffile{pomer.eps}
\caption{Variation of the renormalized energy
$\varepsilon ({\bf k})$ near the chemical potential in a
particular direction due to increasing $(1 \rightarrow 3)$
attractive attractions with an energy cut-off $\Omega_0$. Curve
labelled (4) is for $F_1^s$ attractive beyond the L-P limit and
therefore displays unstable negative dispersion at the
Fermi-surface.} \label{fig:1.}
\end{figure}




   The  cure to the L-P instability is to eliminate
   the region of negative dispersion in fig. (1) through an ansatz for the form of the
   self-energy, to be tested through solution of Eq. (9),
   which is a continuation
   of the form of the self-energy as the instability is
   approached. This choice is unambiguous,
\begin{eqnarray}
  E^{>}_{{\bf{k}}}& = & \epsilon_{\bf{k}} + D(\hat{k}_f),
  E^{<}_{{\bf{k}}} = \epsilon_{\bf{k}} -
  D(\hat{k}_f),
~~for~~|E_{{\bf k}}-\mu| \lesssim D(\hat{k}_f)  \nonumber \\
E_{{\bf k}}& = & \epsilon_{\bf k}~~for~~|E_{{\bf k},\sigma}-\mu|
\gtrsim \Omega_c.
\end{eqnarray}
 In Eq. (13) $D(\hat{k}_f)\geq 0$ all around the Fermi-surface, so that the
 ansatz,
 produces a (in general anisotropic) gap at the chemical
 potential. This ansatz used in Eq. (9) is sufficient  to determine the transition
 temperature $T_g$ and the magnitude and angle dependence of $D(\hat{k}_f)$ just below
 it, as shown below.

   The alternative of deforming the Fermi-surface asks
   for a self-energy $D({\bf k})$ which does not change sign at the chemical potential as one moves
    perpendicular to the
   Fermi-surface but changes sign on going around the
   Fermi-surface. This would require the ansatz
     \begin{eqnarray}
   D({\bf k})=D v_f \delta k_F(\theta),
   \end{eqnarray}
    where
   $\delta k_F(\theta)$ is the change in the Fermi-radius at $\theta$, which one may seek to
   determine from the same self-consistency equation, viz (9). This ansatz is a continuation for the case that as the instability is approached,
   \begin{eqnarray}
    v_{F\ell}^{-1}\partial \Sigma_(\omega,|\bf{q}|) /\partial|\bf{q}||_{0,\bf{k}_F} \rightarrow -1,
    \end{eqnarray}
    This  corresponds to $F_{1s} \rightarrow \infty$ 
in TI problems but even more generally to both $F_{0s},F_{0a}\rightarrow \infty$.
The Mott or Wigner transition is a competing (first order) transition in this case.

Consider next  the $\ell=0$ instabilities or the divergence of the
compressibility or magnetic susceptibility:
   \begin{eqnarray}
   \frac{\kappa}{\kappa_0} = \frac{m^*/m}{1+F_0^s},~~~\frac{\chi}{\chi_0} =
   \frac{m^*/m}{1+F_0^a}.
   \end{eqnarray}
   By employing Ward-identities \cite {Nozieres}, one can show
   that
   \begin{eqnarray}
   \frac{\kappa}{\kappa_0}& = & (1+\frac{d\Sigma}{d\mu})(1+(v_{f0})^{-1}\partial \Sigma/\partial {\bf
   k}|_{k_F})^{-1}, \\
   \frac{\chi}{\chi_0}& = & \left(1+{\bf \sigma}\cdot\frac{d\Sigma}{d(g\mu_B{\bf H})}\right)(1+(v_{f0})^{-1}\partial \Sigma/\partial {\bf
   k}|_{k_F})^{-1}.
\end{eqnarray}
 If the compressibility diverges without the effective mass diverging
or independently of it, the L-P instability requires
$\frac{d\Sigma}{d\mu} \rightarrow \infty$. The ansatz for $D$ to
cure the instability is then again that it is discontinuous at the
chemical potential just as in Eq. (13) leading  to an isotropic
gap in the single-particle excitation spectra in the stable state.
In real physical systems, compressibility of the electronic
system, ignoring its coupling to the lattice is generally not
relevant.

The $\ell = 0$ instability in the antisymmetric channel is
of-course the prelude to ferromagnetism, which is signaled by
$F_0^a \rightarrow -1$ or equivalently through Eq. (17) by ${\bf
\sigma}\cdot\frac{d\Sigma}{d{\bf H}} \rightarrow \infty$. In this
case the ansatz to be made in Eq. (10) is just the
 traditional ansatz in the mean-field
theory of itinerant ferromagnetic instability, viz.
\begin{eqnarray}
E_{{\bf k},\sigma}= \epsilon_{\bf k} +{\bf \sigma}D(\hat{k}_f)
~~for~~|E_{{\bf k},\sigma}-\mu| \lesssim D(\hat{k}_f)  \nonumber
\\
E_{{\bf k},\sigma}= \epsilon_{\bf k}~~for~~|E_{{\bf
k},\sigma}-\mu| \gtrsim \Omega_c.
\end{eqnarray}
The adjustment of the up and down Fermi-surfaces so that the
 chemical potential for the up and down spins is the same is an
 additional condition which must be imposed here as in the
 traditional theory.

For higher $\ell$,  and for Fermions on a lattice, absent the Ward identities,  
only approximate derivation of  the cure to the long wavelength instabilities are possible. (This is also  the case of the familiar finite $\bf {Q}$ instabilities such as CDW or SDW.) One can rely on the fact that Eq. (3) can also be obtained in approximations such as the RPA or generalized RPA, which use approximate interaction functions which we will continue to denote by $F({\bf{ k,k',q}})$ and the approximate single-particle energies. The procedure to be followed is :  Write the interaction in separable form in the irreducible representations (IR's) of the lattice. Check through solution of Eq. (3) or by some equivalent method if the interactions drives a mode in some IR unstable for some value of the coupling constant.  Calculate the single-particle self energy near that point in a consistent approximation. On a lattice the self-energy  may be expanded in irreducible representations of the lattice. ( I will continue to enumerate the IR's by $\ell$ and denote their basis by ${\cal P}_{\ell}$ .) 
\begin{eqnarray}
   \Sigma({\bf k},\omega) = \sum_{\ell} \Sigma_{\ell}(
   |{\bf k}|,\omega){\cal P}_{\ell}(\cos \theta),
   \end{eqnarray}
 For spin-symmetric instabilities, suppose that for {\it any} $\ell$,
   \begin{eqnarray}
   v_{F}^{-1}\partial \Sigma_{\ell}(\omega,|\bf{q}|) /\partial|\bf{q}||_{0,\bf{k}_F} \rightarrow \infty ,~~or
   \end{eqnarray}
  \begin{eqnarray}
   v_{F}^{-1}\partial \Sigma_{\ell}(\omega,|\bf{q}|) /\partial|\bf{q}||_{0,\bf{k}_F} \rightarrow -1 .
   \end{eqnarray}
    The  single-particle energy is as exhibited in fig. (1) for the first case and has the  opposite sign for the second case.  The cure to the instability for the first case is found by the ansatz of the form of Eq. (13).  The cure for the second case is found with the ansatz of the form of Eq. (14). The dependence of the functions $D(\bf{k})$ 
in either case is found from solving Eq. (10) as done below. An explicit example of this approach
including the derivation of (21)for a specific case is given at the end.
   
   For the spin-antisymmetric instabilities, 
 the RPA type equation show only one route to the cure. 
Suppose as the instability is approached,
    \begin{eqnarray}
   {\bf\sigma}\cdot\frac{d\Sigma_{\ell}}{g\mu_Bd{\bf H}} \rightarrow \infty 
   \end{eqnarray}
  for some $\ell$. The cure  is through an ansatz of the form (19). 
  In the absence of spin-orbit scattering spin may be quantized by helicity. In that case or with spin-orbit
 scattering in a crystal with inversion symmetry, no net moment is implied in any
 $\ell$ other than the $\ell=0$ case by (19). The general solution
 in such cases has a gap function of the form $D_{ij}(\hat{k_f})\sigma_j$.
 The Magnetizaion direction rotates around the fermi-surface in the new state 
and there is an anisotropic gap single particle excitations with a given spin or for spin-flip excitations
  
 The relation of these instabilities on a lattice to the L-P instability except for the $\ell=0$ or $1$ cases is
 only conjectural. The instabilities are related to the divergence or zeroes of the eigenvalues of the generalized compressibilities 
 $\partial ^2 F/\partial u_{\alpha}\partial u_{\beta}$ where $u_{\alpha}$ is the strain in the $\alpha$-direction or the generalized magnetic susceptibilities $\partial ^2 F/\partial H_{\alpha}\partial H_{\beta}$, as are the L-P instabilities. 

  I now return to the integral equation for $D$, Eq. (10) with the ansatz of Eq. (13) for
 arbitrary
$\ell$ for the TI or for the lattice case. Eq. (10) may be rewritten as,
 \begin{eqnarray}
1 \approx -F_{\ell}^s\nu(0)^{-1}\int d{\theta}\sin(\theta)\frac{{\cal P}_{\ell}^2(\cos\theta)}{
 D(\theta)} \int_{-\Omega_c}^{\Omega_c}
  d\epsilon \nu(\epsilon) \frac{tanh(\beta D)} {1+cosh(\beta \epsilon)
  sech(\beta D)}.
  \end{eqnarray}

The transition temperature to the new state is then calculated
  from Eq. (24) to be
  \begin{eqnarray}
  T_g \approx \frac{\Omega_c/2}{\ln\left(\frac{F_{\ell}^ s/(2\ell+1)}{1+F_{\ell}^
  s/(2\ell+1)}\right)}.
\end{eqnarray}
For $T$ below $T_g$ and $(T_g-T)/T_g \ll 1$, Eq. (24) also gives that
\begin{eqnarray}
\int {d{\theta}\sin\theta {\cal P}_{\ell}^2(\cos\theta)
 D^2(\theta)} = 6(T_g-T)\Omega_c |F_{\ell}^s|.
 \end{eqnarray}
The requirement that $D(\theta)>0$ in Eq. (23) leads to
\begin{eqnarray}
 D(\theta) =D_0 {\cal P}_{\ell}^2(\cos\theta),
 \end{eqnarray}
 where $D_0$ may be determined from Eq. (26).
(27) is not the only solution of the equation (26).
 Although all solutions with the ansatz (13) have an anisotropic
  gap at the chemical potential, the solution
which minimizes the energy can be easily calculated to be the one
 in which the gap matches the kernel of the integral Eq. (26)
most closely, as (27)does.

 Eq. (27)
 changes the one-particle density of states in a range of
$O(D_0)$ near the chemical potential $\mu$.
 $\mu$ must be determined anew to keep the number of particles
 conserved. For small $D_0\nu(0)$, the new density of states for any $\ell$ is
 \begin{eqnarray}
 \tilde{\nu}(E) \propto \nu(0) (E/D_0)^{1/2} ~~for~~ E\ll D_0
 \end{eqnarray}
The specific heat in this
 state varies at low temperatures as $T^{3/2}$ and the magnetic
 susceptibility as $T^{1/2}$.

 Similarly the spin-antisymmetric instabilities give rise to a gap
 for spin-flip excitations
 for any $\ell$ with the value of the gap varying as
 ${\cal P}_{\ell}^2(\cos\theta)$.

Landau or Pomeranchuk did not consider a cure for their
instabilities. The symmetric spin-channel has been recently
considered \cite{kivelson,metzner} where the stable state is
assumed to have a distortion of the Fermi-surface without a gap. A
L-P  instability for mesoscopic systems also has
been proposed \cite{murthy}. The solution similar to that here was
proposed \cite{cmv} in connection with the special case of the
pseudogap state of the cuprates.  Several
analytical-numerical calculations \cite{metzner,wegner} indicate
a L-P instability in the "d-wave" channel in the
Hubbard model. In analytic calculations collective
spin-fluctuation modes have been argued to lead to negative Landau
parameters \cite{morita}

 {\it Derivation of Eq. (9) and of the form of the self-energy for a specific model of interactions}:
 Consider bosons with bare energy $\Omega_{\bf q}$ interacting
with the fermions through coupling
 functions $g({\bf k},{\bf k+q})$,
 \begin{eqnarray}
 H_{bosons} = \sum_{{\bf k}} \Omega_{{\bf k}}b^+_{{\bf k}}b_{{\bf
 k}} +
 \sum_{{\bf k},{\bf q}}g({\bf k},{\bf k+q})c^+_{{\bf
 k+q}}c_{{\bf k}}(b_{{\bf q}}+b^+_{-{\bf q}}).
 \end{eqnarray}
 The nature of these Bosons is unspecified but it is necessary
 that $lim {\bf q}\rightarrow 0 g({\bf k},{\bf k+q})  \neq 0.$
 Let the boson be unstable in the $lim~ {\bf q}\rightarrow 0$ due to the interaction with the fermions. The
 instability is cured by having a finite expectation values for
 $b_{{\bf q}},b^+_{{\bf q}}$ in the limit $ {\bf q}\rightarrow 0$,
 which may be found from Eq. (9),
 \begin{equation}
 \langle b_{{\bf q}}\rangle = -\sum'_{{\bf k}} \frac{g({\bf k},{\bf
 k+q})}{2\Omega_{{\bf q}}}\langle c^+_{{\bf
 k+q}}c_{{\bf k}}\rangle
 \end{equation}
 The effective {\it mean-field} hamiltonian for the electrons is
 now
 \begin{eqnarray}
 H_{eff} & = &\sum_{{\bf k}}\epsilon_{{\bf k}}c^+_{{\bf k}}c_{{\bf
 k}} + \sum_{{\bf q}}V({\bf k,k+q})c^+_{{\bf
 k+q}}c_{{\bf k}}, \\
V({\bf k,k+q}) & \equiv & g({\bf k},{\bf k+q})(\langle b_{{\bf
q}}\rangle +\langle b^+_{-{\bf q}}\rangle).
\end{eqnarray}
The one-particle Green's function for this hamiltonian in the
self-consistent Brillouin-Wigner approximation is
\begin{eqnarray}
G({\bf k},\omega) = \frac{1}{\omega -\epsilon_{{\bf k}}- D({\bf
k},\omega)},
\end{eqnarray}
 where at $\omega = E_{{\bf k}}\equiv \epsilon_{{\bf k}}+ D({\bf
k},E_{{\bf k}})$,
\begin{eqnarray}
D({\bf k},E_{{\bf k}}) = \sum_{{\bf q}} \frac {(f(E_{{\bf
k+q}})-f(E_{{\bf k}}))|V({\bf k,k+q})|^2}{E_{{\bf k+q}} - E_{{\bf
k}}} = \sum_{{\bf q}}V({\bf k,k+q})\langle c^+_{{\bf
 k+q}}c_{{\bf k}}\rangle.
\end{eqnarray}
The second equality in (31) follows from Eq.(21).
Therefore,
\begin{eqnarray}
\langle c^+_{{\bf
 k+q}}c_{{\bf k}}\rangle = \frac{g({\bf k},{\bf k+q})(f(E_{{\bf k+q}})-f(E_{{\bf
 k}}))(\langle b_{{\bf q}}\rangle+\langle b^+_{-{\bf q}}\rangle}{E_{{\bf k+q}} - E_{{\bf k}}}
 \end{eqnarray}
 Inserting $\langle b_{{\bf q}}\rangle,\langle b^+_{-{\bf q}}\rangle$ from Eq. (30),
the self-consistency Eq. (9) is
 obtained with the Landau-function $F({\bf k,k'})$ identified as
 $\lim {\bf q}\rightarrow 0 [-g({\bf k,k+q})g({\bf k',k'+q})/\Omega_{\bf q}]$.
 Similar results can also be obtained from a purely electron-electron interaction
 model in which
$(b_{{\bf q}}+b^+_{-{\bf q}})$ in (26) is repalced by $c^+_{{\bf
 k'+q}}c_{{\bf k'}}$ and $g(\bf{k,k+q})$ by a function
 $V(\bf{k,k',q})$.In that case both the direct and the exchange interactions should be considered.
 
 If the Boson is a fluctuation of a non-conserved quantity, its propagator near the instability is allowed  the form, 
 \begin{eqnarray}
  (i\omega +a^2 q^2+\epsilon)^{-1}. 
  \end{eqnarray}
  $\epsilon$ is the distance to the instability as a function, say of 
the coupling constants g at T=0. The fermion self-energy at the chemical
 potential  for small deviation $\delta q$ from ${\bf k}_{F}$ normal to 
the fermi-surface  is calculated to be  $\propto 
 |g(\hat{k}_F,\hat{k}_F)|^2 \delta q\ln(q_c/(\delta q+\epsilon)$ in 3 
dimensions and  $\propto  
|g(\hat{k}_F,\hat{k}_F)|^2 (\delta q+\epsilon)^{1/2}\ln(q_c/\delta q)$ in 2
 dimensions.(The singularities at ${\bf k}={\bf k}_F$ as a function of energy are
weaker.) In either case the derivative of the self-energy satisfies  
(21) as the instability is approached. The cure to the instability is 
therefore a state with an anisotropic gap at the chemical potential. 
This singularity in self-energy is reminiscent of the 
 Hartree-Fock sigularity due to (unscreened) coulomb interactions.
 That singularities of-course disappears due to screening. 
Near an instability with a propagator of the form (36),
 the generated long-range interaction is protected.

 I wish to acknowledge useful discussions with E.Abrahams,B.I.Halperin,
 H.Maebashi, K.
 Miyake, G.Murthy, R.Shankar and P.Woelfle.

\end{document}